\documentclass[fleqn,twoside]{article}
\usepackage{espcrc2}

\usepackage{graphicx}
\usepackage[figuresright]{rotating}
\usepackage{epsfig}

\newcommand{\ppbar}{p\overline{p}}

\newcommand{\AmS}{{\protect\the\textfont2
  A\kern-.1667em\lower.5ex\hbox{M}\kern-.125emS}}

\title{Design and Expected Performance of the BTeV RICH}

\author{Steven. R. Blusk \newline
Syracuse University \newline
for the BTeV Collaboration}

\begin{document}

\begin{abstract}
	The BTeV experiment is a $b$-physics experiment
designed to conduct precision tests of the CKM description of CP violation and
study rare processes involving bottom and charm hadrons. The experiment
will be located in the C0 interaction region at the Fermilab Tevatron, and
is intended to begin data-taking around 2007-2008. One of the most important 
elements of the BTeV spectrometer is the ring-imaging Cerenkov detector 
(RICH) which is used for particle identification. In this article we describe
the BTeV RICH and present its expected performance.
\vspace{1pc}
\end{abstract}

\maketitle

\section{Introduction}

   The BTeV experiment is designed to carry out a comprehensive program of
precision measurements of the Cabibbo-Kobayashi-Maskawa (CKM) matrix as they 
relate to CP violation in the b-quark sector. In addition, the experiment
will conduct a wide array of measurements on rare decays in both the bottom
and charm quark sector. These measurements will provide perhaps the most 
stringent tests of the CKM description of CP violation as well as provide 
complementary probes to the LHC for the presence of physics which is not
described by the standard model (``new physics''). 

   The first indication of a non-zero value for $\sin(2\beta)$ as measured in
$B\to J/\psi K_S$ decays was presented by CDF~\cite{cdf_sin2b}. The B-factory experiments,
BaBar and Belle, which operate on the $\Upsilon (4S)$ have improved significantly on this 
measurement, and now have unambiguous confirmation of a non-zero value for 
$\sin(2\beta)$~\cite{sin2b}. Despite their impressive achievements, 
construction of a dedicated $b$-physics experiment
at a hadron collider holds several key advantages. First, the production cross-section is
$\approx$10,000 times larger at the Tevatron than at the $\Upsilon (4S)$. Secondly, $e^+e^-$ machines operating at the $\Upsilon (4S)$ cannot perform CP violation measurements in the $B_s$ system, whose decays
can be used to probe the angles $\gamma$ and $\chi$. Given these advantages,
$B$-factories operating at the $\Upsilon (4S)$ (at a luminosity of 
$10^{34}$ cm$^{-2}$ s$^{-1}$) cannot compete with a hadron collider, even on the time scale of 2007.

   In $\ppbar$ collisions at $\sqrt{s}$=2 TeV, $b$ quark pairs
exhibit a high degree of correlation. In particular, when one $b$ quark has large
pseudorapidity ($\eta$), the other $b$-quark is also preferentially produced at
large $\eta$. As a result, both $b$-quarks are often either both forward, or both 
backward, with respect to the direction of the proton beam. The advantages of this 
are two fold. First, at large $\eta$, the $b$-hadron is boosted with $< \beta\gamma >\sim 6$, which results in larger decay lengths. These larger decay lengths aid in
separating the $b$-daughter particles from those in the interaction vertex.
Not only does this improve the overall detection efficiency, but it allows BTeV to
trigger on $b$-hadron decays at the lowest level of the trigger.

   Unlike CDF~\cite{cdf} and D0~\cite{d0} which are central detectors, BTeV~\cite{btev} is a forward detector optimized for $B$ physics. The detector consists of an array of 31 pixel modules 
inside a dipole magnet centered on the interaction region. Tracking beyond the
pixel system is provided by 7 tracking stations which are each a combination of silicon-strip detectors for the forward high-occupancy region and straw-tube drift chambers
at larger angles. Downstream of the last tracking chamber is a PbWO$_4$ 
electromagnetic calorimeter and a muon detection system. The RICH is situated 
between the sixth and seventh straw/silicon tracking chambers. 

	The RICH is used for separating charge particle species from one another.
When a particle exceeds the speed of light in a dielectric medium, Cerenkov photons
are emitted at a characteristic angle (called the Cerenkov angle) given by $\cos\theta_c=1/\beta n$, where $\beta=v/c$ is the speed of the particle relative to the speed of light in the radiating medium, $n$ is the index
of refraction of the medium and $\theta_c$ is the Cerenkov angle. By using the momentum
measured in the tracking system and measuring the Cerenkov angle, the
particle type (ie., the mass of the radiating particle) can be deduced. 

\section{BTeV RICH}

\begin{figure}[bht]
\vspace{0.0cm}
\centerline{\epsfig{figure=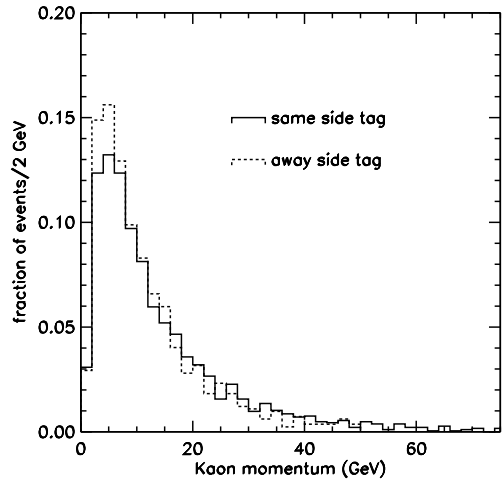,height=3.0in}}
\vspace{-1.0cm}
\caption{\label{kaon_mom_frac} Momentum distribution for kaons which provide a flavor tag for $B$-meson CP eigenstates. Shown are the distributions for kaons which are produced in the fragmentation of the $b$-quark which produces the CP-tagged $B$ meson (same-side tag), and kaons which are produced in the decay of the second $B$-hadron (away-side tag).}
\end{figure}

	To separate kaons from pions in two-body $B$ decays, we require that the RICH
provides at least 3$\sigma$ separation between kaons and pions up to 70 GeV/c. 
At the low momentum end, the RICH is used for flavor tagging. These particles 
peak at lower momentum, as shown in Fig.~\ref{kaon_mom_frac}, and therefore the low momentum cut-off is set by the spectrometer acceptance, which drops off around 2.5 GeV/$c$.

\begin{figure}[bht]
\vspace{0.0cm}
\centerline{\epsfig{figure=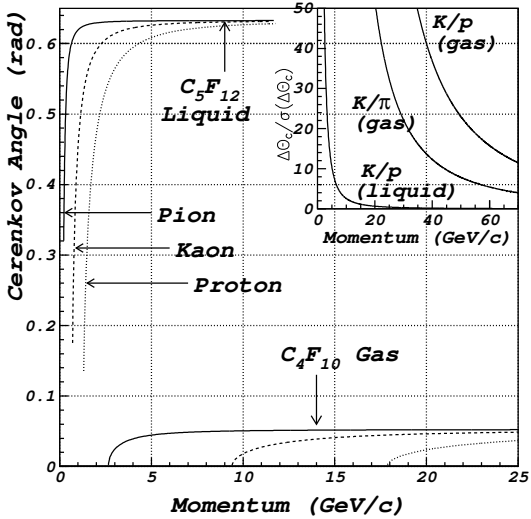,height=3.0in}}
\vspace{-1.0cm}
\caption{\label{thetac} Cerenkov angle as a function of momentum for the gas and liquid radiators for pions (solid), kaons(dashed) and protons (dotted). The inset shows the separation power as a function of momentum.}
\end{figure}

	To achieve particle separation over this large momentum range a heavy gas is
required. We chose $C_4 F_{10}$ because of it is among the heaviest gases which has
high transparency with respect to visible light and is a gas at room temperature.
This gas has also been used in a number of other experiments~\cite{delphi,hera_b,hermes}
as well as LHCb~\cite{lhcb_rich}. No gas can provide kaon/proton separation below
about 9 GeV, and we therefore are designing a 1 cm thick $C_5 F_{12}$ liquid radiator system. The Cerenkov angle as a function of momentum for the two radiators are
shown in Fig.~\ref{thetac}. The gas radiator has n=1.00138, whereas the liquid has 
n=1.24, resulting in vastly different Cerenkov angles and hence momentum coverage.

\begin{figure}[bht]
\vspace{0.0cm}
\centerline{\epsfig{figure=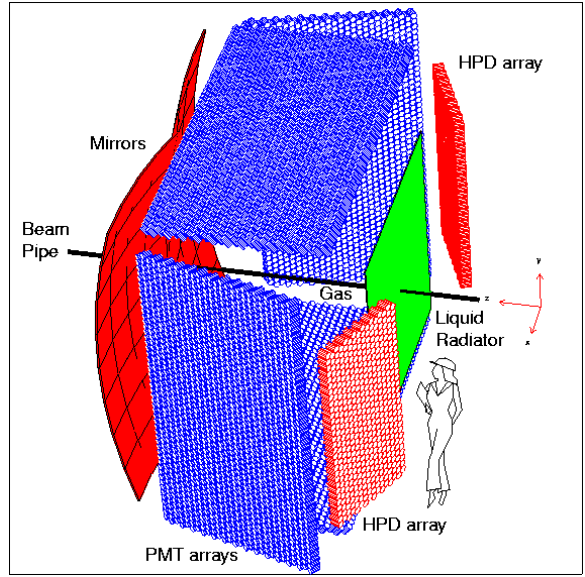,height=3.0in}}
\vspace{-1.0cm}
\caption{\label{btev_rich} Cartoon showing the conceptual design of the BTeV RICH.}
\end{figure}

	A cartoon of the baseline RICH for BTeV is shown in Fig.~\ref{btev_rich}. The system consists of a 1 cm $C_5 F_{12}$ liquid radiator, followed by $\sim$3 m of $C_4 F_10$ gas. At the back end of the RICH, two spherical mirrors, each tilted by 15$^o$, focus Cerenkov photons from the gas radiator onto one of two arrays of hybrid photodiodes (HPDs) located on either side of the BTeV dipole magnet. The arrays consist of about 1000 HPD's, each with 163 channels. Photons from the liquid radiator are detected directly by 
arrays of photomultiplier tubes which cover the walls of
the RICH vessel and are tilted at $60^o$ so that photons strike perpendicularly, on average. A total of about 5000 PMTs are required. Because of the large difference in Cerenkov angle between the gas and liquid radiator, gas photons are detected almost entirely in the  HPD's and liquid photons in the PMT's.
	
	At 70 GeV/c, the difference in $\theta_c$ between kaon and pions is only 0.4 mrad,
and therefore, to achieve at least 3 $\sigma$ separation, we must keep the
total uncertainty in the Cerenkov angle ($\sigma$) below about 0.130 mrad. Detailed simulations have been carried out, and they indicate that we expect to detect $\sim 80$ photons per track. After removing photons which are consistent with more than one track, we expect about 60-65 photons per track on average. This implies that the single photon angular resolution must be kept below 1 mrad. The primary contributions to the intrinsic angular resolution are from  photon position resolution, emission point uncertainty and chromatic uncertainty.
In collaboration with DEP we have developed a 8.3 cm diameter, 163 channel HPD which 
will contribute an uncertainty of 0.51 mrad to the Cerenkov angle uncertainty. The
emission point uncertainty contributes another 0.51 mrad uncertainty. The chromatic
uncertainty yields a shallow minimum at 0.37 mrad if we restrict our wavelength detection
to larger than 280 nm. All together, we expect to achieve $\sim$0.8 mrad uncertainty
on the Cerenkov angle per photon, or $\sim 100 \mu$rad per track. The expected separation (in number of sigma's) between kaons and pions as a function of momentum is shown in the inset in Fig.~\ref{thetac}.
	
	The liquid radiator extends the coverage for kaon/proton
separation down to the low momentum cutoff of the BTeV spectrometer ($\sim$2 GeV/c).
For the liquid radiator, the difference in kaon-proton Cerenkov angle is
about 5.3 mrad at 9 GeV. We therefore are required to keep the Cerenkov
angle resolution below about 1.7 mrad per track. To determine the expected
number of detected photons, we simulated a photodection
system consisting of large arrays of single anode PMTs along the walls of 
the RICH tilted at an angle of 60$^o$. We also included the shadowing produced
by mu-metal tubes which will likely be necessary to cope with the residual magnetic
field of the interaction region magnet. We simulated both 2 in. and 3 in.
diameter tubes. The 3 in. diameter tubes provide sufficient resolution
and are more cost effective. The simulation shows that for a 3 in.
diameter PMT, we would expect to detect about 15 photoelectrons, and, with this 
granularity, the Cerenkov angle uncertainty per photon is about 6.5 mrad. This dominates
over the chromatic and emission point uncertainty, and therefore the 3 inch tube
is our baseline choice for detecting photons from the liquid radiator. We expect to obtain a total uncertainty in the Cerenkov angle of 1.7 mrad per track.

	The expected angular uncertainties for the gas and liquid radiator systems are summarized in Table~\ref{angular_resolution}.

\begin{table}[htb]
\caption{The expected Cernkov angle uncertainties from various sources and the total uncertainty for the gas and liquid RICH systems.}
\label{angular_resolution}
\begin{tabular}{@{}ccc}
\hline
  		   &     Gas  &     Liquid      \\
Source	   &   System   &    System    \\
		   &   (mrad)   &   (mrad)      \\
\hline 	
Segmentation   &     0.51   &     5.3       \\
Emission Point &     0.51   &     0.4      \\
Chromatic Uncertainty & 0.37 &    3.7      \\
\hline
Total per photon   &   0.81   &     6.5       \\
\hline
Total per track   &    0.105   &    1.7       \\
\hline
\end{tabular}
\end{table}

	Photo-electrons (PE) produced from the HPD photo-cathode are accelerated 
by a 20 kV potential and impinge upon a silicon-pixel array. The PEs are cross-focused and demagnified by a factor of four using two focusing electrodes operating at 19.89 and 15.6 kV. The photoelectrons deposit all of their energy in the silicon pixel, giving rise to a signal of 5000 electrons. The HPDs will be read out with a custom ASIC
($VA\_ BTeV$) being developed in
collaboration with IDE AS Norway. The ASIC is a 64-channel circuit featuring a low noise front 
end amplifier capable of achieving a noise of $\sim$700 electrons, a shaper and a comparator.
Thresholds for each channel can be fine-tuned via an initialization sequence. Each channel has 
binary output (ON or OFF). To reach this low noise level, the front end is not clocked but rather 
is always active after it is initialized. One result of this feature is that after a channel is
hit, it requires about 200 ns for the front end to return to its baseline level. Since the Tevatron
will have a beam crossing every 132 ns, a hit channel is unavailable for the next beam crossing. The HPD pixel occupancy was simulated at an instantaneous luminosity of 
$2x10^{32}$ cm$^{-2}$ s$^{-1}$ and the results show that the number of channels per HPD with hits in two consecutive beam crossings is less than 10\% in the busiest HPD.
In the busiest 12 HPDs, the rate of consecutive hits is approximately 2-3\% and outside this region it is well below 1\%. We therefore are confident that the occupancy is low enough that inefficiency due to hits in consecutive beam crossings is a non-issue. A more detailed report on the performance of the HPD's is given in the references~\cite{raym}.
The PMTs will most likely be read out with a similar chip to the ${\rm VA\_ BTeV}$, modified to deal with the larger PMT signal. 

\section{Expected Performance}

\begin{figure}[bht]
\vspace{0.0cm}
\centerline{\epsfig{figure= 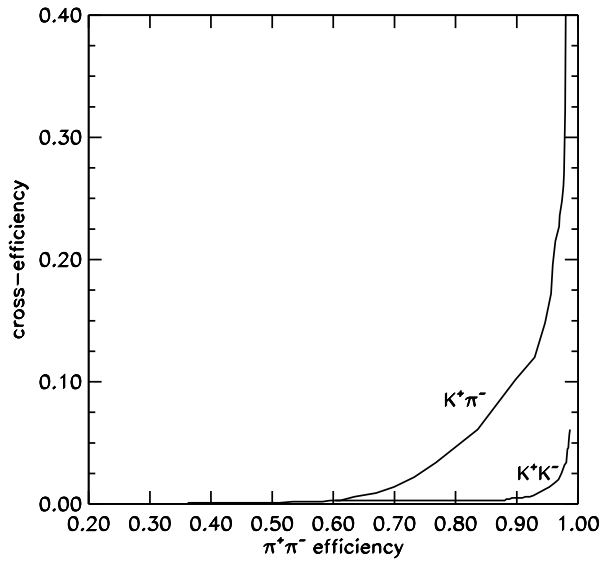,height=3.5in}}
\vspace{-1.0cm}
\caption{\label{eff_vs_fake_pipi_2minbias} Cross-efficiency for identifying a $B^0\to\pi^+\pi^-$ event as a $B^0\to K^+\pi^-$ or a $B^0\to K^+K^-$ event as a function of the tagging efficiency for $B^0\to\pi^+\pi^-$ using the gas portion of the RICH.}
\end{figure}

	We have simulated both the gas and liquid radiator systems. One measure of the
performance of the gas system is to compare the efficiency for tagging both pions in 
$B\to\pi^+\pi^-$ versus the cross-efficiency for assigning one or more of the of the pions 
to be a kaon. The results of the simulation are shown in Fig.~\ref{eff_vs_fake_pipi_2minbias}.
The study indicates that we can achieve a $\pi^+\pi^-$ efficiency of 80\% (90\%) while 
keeping the $K^-\pi^+$ misidentification rate at 5\% (12\%).

Since the separation between pions and kaons increases as the number of B-daughter tracks increase (lower average momentum per track), it is clear that the gas portion of the BTeV RICH will provide excellent $K/ \pi$ separation. 

\begin{figure}[ht]
\vspace{2.0cm}
\centerline{\epsfig{figure= 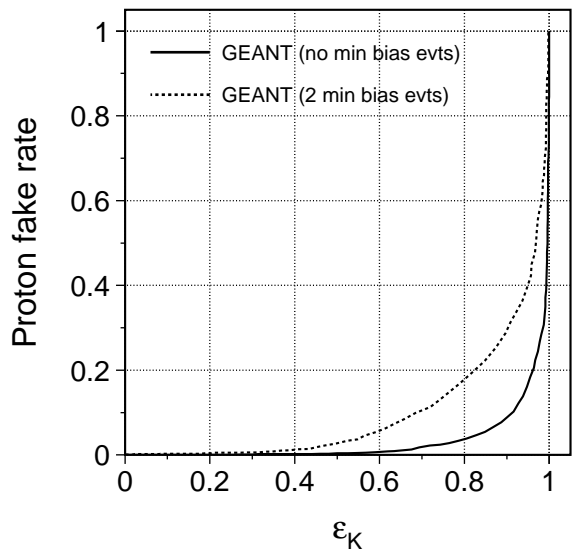,height=3.25in}}
\vspace{-3.0cm}
\caption{\label{eff_vs_fake_kp} Probability for misassigning a proton to be a kaon (proton fake rate) as a function of the kaon tagging efficiency ($\epsilon$) for the liquid portion of the RICH.}
\end{figure}

The liquid radiator system was also simulated using $b\overline{b}$ events. The resulting proton fake rate versus kaon efficiency is shown in Fig.~\ref{eff_vs_fake_kp}. The analysis was performed using a full GEANT simulation with 0 and 2 (Poisson distributed) minimum-bias events added into each event. We conclude that we can achieve $\sim$80\% kaon efficiency while rejecting $\sim$85\% of the protons. A study to assess the impact on flavor tagging was conducted, and our simulations indicated that this system can improve the flavor tagging for $B_s$ by 25\% and by 10\% for $B^0$. This improvement comes mainly from improved rejection of protons faking kaons below 9 GeV/c, where neither the kaon nor proton produce Cerenkov radiation in the gas. The significantly larger improvement in $B_s$ over $B^0$ is attributed to the large flavor tagging efficiency for same-side kaon tags over same-side pion tags.

	In summary, we plan to construct a RICH detector which includes both gas and liquid radiator subsystems. The gas radiator system uses a $C_4 F_10$ gas radiator and HPDs for photodetection. The gas system provides $\ge$4$\sigma$ $K/ \pi$ separation for momenta in the range from 3-70 GeV/$c$ and $\ge$10$\sigma$ $p/K$ separation for momenta greater than 9.5 GeV/$c$. The liquid radiator system utilizes a 1 cm thick $C_5 F_{12}$ liquid radiator whose Cerenkov photons are detected in an array of 3 in. PMT's which line the walls of the gas vessel. The liquid system extends the capabilities of the RICH to provide excellent $K/p$ separation for momentum below 9 GeV/$c$ and will improve BTeV's flavor tagging capabilities.

\end{document}